\documentclass{format170x240multiauthor}
\usepackage[T1]{fontenc}
\usepackage{makeidx}
   \makeindex
\usepackage{amsmath}
\usepackage{textcomp}
\usepackage{graphicx}
\usepackage{cite}
\usepackage{url}
\begin{document}

\pagenumbering{arabic}

\chapter{Stochastic evolutionary game dynamics}
\label{ch1}
\chapterauthor[]{Arne Traulsen and Christoph Hauert}

\vspace{-5cm}
{\sffamily To appear in  {\bf ``Reviews of Nonlinear Dynamics and Complexity''  Vol. II}, Wiley-VCH, 2009, edited by H.-G. Schuster }
\vspace{6cm}

\section{ Game theory and evolution}

Modern game theory goes back to a series of papers by the mathematician John von Neumann in the 1920s. This program started a completely new branch of social sciences and applied mathematics. 
 This early work on game theory is summarized in the seminal book ``The Theory of Games and Economic Behavior'' by John von Neumann and Oskar Morgenstern \cite{neumann:1944ef}. Initially, game theory was primarily focused on cooperative game theory, which analyzes optimal strategies assuming that individuals stick to previous agreements. 
In the 1950's, the focus shifted to non-cooperative games in which individuals act selfish to get the most out of an interaction. At that time, game theory had matured from a theoretical concept to a scientific field influencing political decision making, mainly in the context of the arms race of the cold war. 

The basic assumption was that individuals act rationally and take into account that their interaction partners know that their decisions are rational and vice versa. Based on a common utility function that individuals maximize, the actions of others can be predicted and the optimal strategy can be chosen. However, the underlying assumption of rationality is often unrealistic. Even in simple interactions between two individuals $A$ and $B$, it is difficult to imagine fully rational decision making,
as this often leads to an infinite iteration: $A$ thinks of $B$, who is thinking of $A$, who is thinking of $B$ and so on.  

One way to avoid this situation in economy is the idea of bounded rationality \cite{selten:1990le,gigerenzer:2002aa}. If the cost of acquiring and processing information is taken into account, individuals can no longer be assumed to do a fully rational analysis of the situation. Instead, their rationality is bounded and the way they make decisions is very different. 
In this case the prediction of the action of others is usually no longer possible. Nonetheless, bounded rationality turned out to be a crucial concept to overcome the problematic rationality assumption in economics. 

A completely different way to deal with this problem in biology has been proposed by John Maynard Smith and George Price in the 1970s \cite{maynard-smith:1973to,maynard-smith:1982to}. They introduced the idea of evolutionary game theory.  
The utility function (which is usually the payoff from the game) is identified with the reproductive fitness. In this way, successful strategies spread in the population and less successful strategies diminish. Two important aspects differ from the traditional setting of classical game theory: 
(i) Evolutionary game theory always deals with populations of individuals instead of two (or few) players.
(ii) In contrast to the static classical game theory analysis, evolutionary game theory provides a natural way to introduce a dynamics into the system. 

There are two interpretations of evolutionary game dynamics. The first one is the traditional setting, in which strategies are encoded by the genome of individuals and successful types spread in the population due to their higher reproduction. 
Examples from biology include the competition of different bacterial strains \cite{kerr:2002xg}, cooperation in virus populations \cite{turner:1999hp}, or the cyclic dominance of mating strategies in lizards \cite{sinervo:1996le,sinervo:2006aa}. 
Biological reproduction selects successful strategies and does not require rational agents or other forms of cognitive abilities.

The second interpretation is cultural evolution. In this setting, successful behaviors are copied by other individuals through imitation. Successful strategies 
propagate through imitation and learning. Although individuals now have to make decisions, this is very different from the rational decisions in classical game theory. Instead of analyzing the situation in detail, the players just imitate those that are more successful. Such strategies are possible even with minimal cognitive premises. This approach is taken for comparisons between predictions of evolutionary game theory and behavioral studies. 

More recently, ideas of evolutionary game theory have been reintroduced to economics, where they are tackled with great mathematical rigor \cite{sandholm:2007bo}.

\section{The replicator dynamics}

Traditionally, evolutionary game dynamics is described for very large, unstructured populations. 
In this case, a differential equation governs the evolution of the densities of the different strategies
\cite{zeeman:1980ze,taylor:1978wv,hofbauer:1998mm}, 
\begin{equation}
\dot x_i = x_i \left( \pi_i - \langle \pi \rangle \right).
\label{repdyn}
\end{equation}
Here, $x_i$ is the fraction of type $i$ in the population, $\pi_i$ is the
fitness of this type and $ \langle \pi \rangle $ is the average payoff in
the whole population. If the fitness of a type is above the average fitness in the population, its density will increase. If the fitness is below the average fitness, then
the corresponding density will decrease.  
Each type $i$ has a fixed strategy. 
If the fitness values $\pi_i$ are fixed, we speak of constant selection. 
In general, $\pi_i $ depends on the composition of the population, i.e.\ on the fractions of all other strategies $x_j$. 
Then, $ \langle \pi \rangle $ becomes quadratic in the fractions $x_j$. Therefore, the dynamics is nonlinear in general. Since the variables in the replicator equation represent the fractions of each strategy in the population, the natural coordinate system is a probability simplex, i.e., for 2 strategies, we have a line, for 3 strategies an equilateral triangle, for 4 strategies an equilateral tetrahedron and so on. 

\begin{figure}[t]
\def\capfrac{1}
\begin{center}
\includegraphics[width=1.0\textwidth]{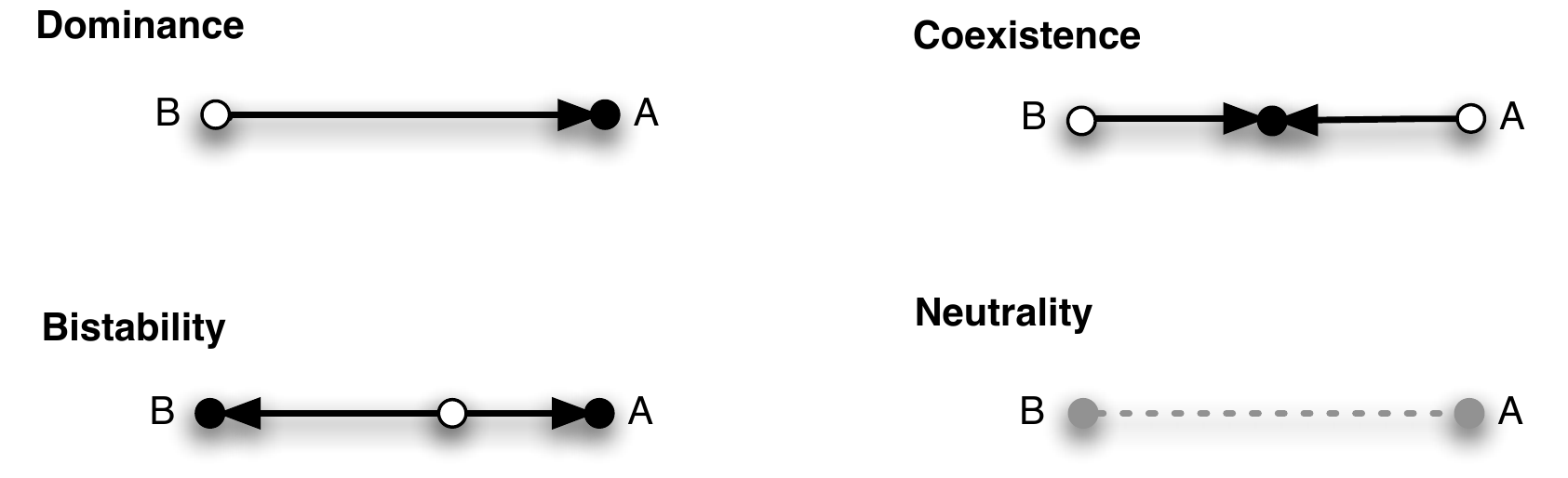}
\end{center}
\Caption{
The four 
dynamical scenarios of evolutionary $2 \times 2$ games. The arrows 
indicate the direction of selection, filled circles are stable fixed points and open circles unstable fixed points. In the neutral case, the entire line consists of neutrally stable fixed points. }
\label{fig1}
\end{figure}

As an example, let us consider the simplest possible games. These are so called two player normal form games or $2 \times 2$ games. Such games can be described by a payoff matrix of the following form
\begin{equation}
\bordermatrix{
  & A & B \cr
A & a & b \cr
B & c & d \cr}.
\end{equation}
This is a way to specify the interactions of two types: If $A$ interacts with another $A$,
it obtains $a$ and $b$ if it interacts with $B$. Similarly, $B$ obtains $c$ from interactions with $A$ and $d$ from interactions with $B$. The payoffs are determined by the fraction of interactions with a given type. Since we have only two types, the population state is fully determined by $x=x_1=1-x_2$. The payoffs are then 
$\pi_A = a \, x+b(1-x)$ 
and
$\pi_B = c \, x+d(1-x)$. This leads to the replicator equation
\begin{equation}
\dot x = x(1-x) \left[ (a-b-c+d) x + b-d \right].
\end{equation}
Apart from the trivial fixed points $x=0$ and $x=1$, the replicator equation can have a third fixed point $x^\ast$
for $a>c$ and $d>b$ or for $a<c$ and $d<b$,
\begin{equation}
x^{\ast} = \frac{d-b}{a-b-c+d}.
\end{equation}
We can distinguish four generic cases \cite{nowak:2004aa}, see Fig.~\ref{fig1}:
\begin{itemize}
\item {\bf Dominance.} 
In this case, one strategy is always a better choice, regardless of the action of the opponent. 
Either $A$ dominates $B$ ($a>c$ and $b>d$) or $B$ dominates $A$ ($a<c$ and $b<d$). In the first case, the fixed point at $x=1$ is stable and the fixed point at $x=0$ is unstable and vice versa in the latter case. 

\item {\bf Bistability.} This is the case for $a>c$ and $d>b$. The fixed points at $x=0$ and $x=1$ are stable, separated by an unstable fixed point $x^{\ast}$. The corresponding games  are called coordination games. 
What is the best strategy in such a game? As a first approximation, one can ask for the maximum payoff in the equilibrium. However, if the opponent is unreliable, one should also try to avoid large losses. This leads to the concept of risk dominance: The strategy that has a larger basin of attraction is called risk dominant. In our case, strategy $A$ is risk dominant for $a+b>c+d$ (or, equivalently $x^{\ast}< \frac{1}{2}$). For $a+b<c+d$ (or $x^{\ast}> \frac{1}{2}$), strategy $B$ is risk dominant. 

\item {\bf Coexistence.} For $a<c$ and $b>d$, there is a stable fixed point at $x^{\ast}$. Hence, the population becomes a stable mixture of $A$ and $B$ types. Both $x=0$ and $x=1$ are unstable fixed points. 

\item {\bf Neutrality.} For $a=c$ and $b=d$, the replicator dynamics predicts neutrally stable fixed points for all values of $x$. While this non-generic case is of limited interest in the replicator dynamics, neutral selection becomes an important reference case in the stochastic evolutionary dynamics of finite populations. 
\end{itemize}

In a genetic setting, the replicator equation is obtained when individuals reproduce at a rate proportional to their fitness. 
In cultural settings, the replicator equation results from individuals imitating better performing actions with a probability proportional to the expected increase in the payoff. 
Moreover, the replicator equations are intimately connected to the Lotka-Voltera equations describing predator-prey dynamics in theoretical ecology. In fact, a direct mathematical mapping from one framework to the other exists \cite{hofbauer:1998mm}.

In general, replicator equations can exhibit very rich dynamics. However, in most cases, the location and stability of fixed points can be determined analytically. In addition, symmetry properties can lead to constants of motion. In this case, the dynamics follows closed orbits. In higher dimensions, the replicator equations can also exhibit deterministic chaos \cite{sato:2002le,sato:2003le,schuster:1995le}.
However, the replicator dynamics is just one particular variant of a deterministic selection dynamics. 
If individuals switch to better strategies independent of the increase in the payoff, one obtains a 
dynamics sometimes called ``imitate the better''. 
Naturally, this would change the form of the differential equation \eqref{repdyn}.
The replicator equation and ``imitate the better'' are two examples of non-innovative selection dynamics, i.e. a strategy that goes extinct will never re-appear. However, in social models, individuals could also choose the ``best reply'' 
to the current composition of the population, even if such a strategy is not present in the population.  This is an example of innovative game dynamics for which equally simple differential equations as \eqref{repdyn} can no longer be 
defined. 

\section{Evolutionary games in finite populations}

While replicator equations have very nice mathematical properties and provide many important insights into evolutionary game dynamics, they essentially describe the deterministic dynamics in an infinitely large population. In general, it is not a priori clear under which circumstances this is a good approximation of the dynamics in a realistic system which is finite and subject to fluctuations. One important source of fluctuations is internal noise arising from the finiteness of the population.  
To answer these questions, we take the following approach: First, we describe evolutionary game dynamics as a stochastic process in a finite population. Then, we increase the population size and infer under which circumstances we recover the replicator dynamics. This also allows us to identify parameter combinations for which it is a good approximation. Moreover, we can 
investigate whether other limits result in qualitatively different dynamics.

\subsection{Stochastic evolutionary game dynamics}

Evolutionary games in finite populations have been considered for a long time
in various fields such as theoretical ecology, behavioral economics or sociology. 
For computer simulations of multi-agent systems, this is the only natural approach to model the dynamics. 
In all cases, we have to specify the microscopic mechanisms which govern 
the transmission of strategies from one individual to another. 
Examples for such selection dynamics are 
\begin{itemize}

\item {\bf Pairwise comparison processes.} 
In this class of processes, two individuals,  a focal individual and a role model, are sampled at random from the population. The focal individual accepts the strategy of the role model with probability $p$, depending on a payoff comparison.
If both individuals have the same payoff, the focal individual randomizes between the two strategies. For example, the probability $p$ could be a linear function of the payoff difference \cite{traulsen:2005hp},
\begin{equation}
p=\frac{1}{2}+w \frac{\pi_f - \pi_r}{\Delta \pi}.
\label{fermieq}
\end{equation}
Here, $w$ ($0 \leq w \leq 1$) is the intensity of selection,
which specifies the importance of neutral drift compared to the selection dynamics.  
This determines the noise intensity.  The functions $\pi_f $ and $ \pi_r$ are the payoffs of the focal individual and the role model, respectively, and $\Delta \pi$ is the maximum payoff difference. For $ w \ll 1$, one speaks of "weak selection". Most importantly, the focal individual does not always switch to the better strategy - sometimes, it also adopts worse strategies. 

 One common choice of a nonlinear function of the payoff difference for $p$ is the Fermi function from statistical mechanics, leading to 
\begin{equation}
p=\left[1+e^{w(\pi_f - \pi_r)} \right]^{-1}.
\label{fermieq}
\end{equation}
Here, the intensity of selection
relates to an inverse temperature and can be any positive number. For weak selection $ w \ll 1$, the probability $p$ reduces to a linear function of the payoff difference. 
For strong selection, $w \to \infty$, this process converges to the imitation dynamics. 
In this case, $p$ becomes a step function being positive for $\pi_r > \pi_f $
and negative for $\pi_r < \pi_f $.
In this limit, only the order of the payoffs is important - an arbitrarily small difference in the payoffs leads to the same outcome.

\item {\bf Moran process.} The Moran process is a classical model of population genetics \cite{moran:1962ef} and has been transferred to game theory only recently \cite{nowak:2004pw,taylor:2004wv}. Here, one individual is chosen at random, but proportional to fitness. 
This individual produces one identical offspring. To keep the number of individuals constant, 
a randomly chosen individual is removed from the population before the offspring is added.
The Moran process represents a simple birth-death process. 
To illustrate the selection mechanism, it is instructive to think of a roulette wheel in which the size of the different fields is proportional to the fitness. Fitness is a convex combination of a background fitness (usually set to one) and the payoff from the game, i.e. $f=1-w+w \pi$. Again, $w$ determines the intensity of selection. For $w=0$, selection is neutral and we have an undirected random walk. For $w \to 1$, fitness equals payoff. However, if the payoffs $\pi$ can become negative, there is a maximum intensity of selection, because the fitness has to be positive. This restriction can be overcome if fitness is defined as an exponential function of payoff, 
$f=\exp \left[ w \pi \right]$. In this case, the selection intensity $w$ can take any positive value \cite{traulsen:2008aa}. 

\item {\bf Wright-Fisher process.} 
The Wright Fisher process has its roots in population genetics as well. 
In contrast to the selection dynamics in the Moran process, where only one individual reproduces at a time, the Wright-Fisher process reflects discrete generations. In every generation, each of the $N$ individuals produce a large number of offspring, proportional to their fitness. From this large offspring pool, a new generation again of size $N$ is randomly sampled. 
Thus, the composition of the population can change much faster. In principle, the population could go back to a single ancestor in a single generation. This reflects the fact that the Wright-Fisher process is not a simple birth-death process, but a more general Markov process. Although it becomes very similar to the Moran process under weak selection, the fixation probabilities (see next section)
cannot be calculated exactly \cite{imhof:2006aa}.

\end{itemize}
The replicator equation determines how the frequencies of strategies in the population change: For each point in state space, the direction of selection and its velocity is determined. 
For finite populations, we have to deal with probabilities instead. Moreover, the state space is discrete. In this discretized space, we can calculate into which direction the system will evolve with what probability. 
This imposes significant restrictions on analytically accessible results. However, the general properties can already be inferred from the simplest possible case, consisting of only two strategies. 

\subsection{Fixation probabilities}

For non-innovative game dynamics, we have two absorbing states: either all individuals are of type $A$ or of type $B$.
One important determinant of the dynamics is given by the probabilities of fixation:
If a mutation leads to a new strategy, how likely is it that this individual takes over the entire population? 

For the sake of simplicity, let us focus on birth-death processes in which only a single individual reproduces at a time (we thus exclude processes as the Wright-Fisher process). We consider a population of size $N$ where the number of $A$ players is $j$ and the number of $B$ players is $N-j$.
The probability to increase the number of $A$ individuals from $j$ to $j+1$ is $T^+_j$. Similarly, $T_j^-$ is the probability to decrease $j$ by 1. Our goal is to calculate the fixation probabilities $\phi_j$, i.e.\ the probability that $j$ individuals of type $A$ succeed and take over the population.
For the absorbing states, we have  
\begin{eqnarray}
\phi_0 = 0 \quad \hbox{and} \quad
\phi_N  =  1.
\end{eqnarray}
For the intermediate states, the fixation probabilities are given by
\begin{equation}
\phi_j = T_j^- \phi_{j-1} + (1-T_j^--T_j^+) \phi_j + T_j^+ \phi_{j+1}.
\end{equation}
We can rearrange this to
\begin{equation}
0 = -T_j^- \underbrace{(\phi_j-\phi_{j-1})}_{y_j} + T_j^+\underbrace{( \phi_{j+1} - \phi_{j})}_{y_{j+1}}.
\end{equation}
This equation can be written as a recursion for the differences between fixation probabilities, 
$y_{j+1} = {\gamma_j} y_j$, where ${\gamma_j} = \frac{T_j^-}{T_j^+} $.
Using this iteration, we find
\begin{eqnarray}
y_1 &=& \phi_1 - \phi_0 = \phi_1\\
y_2 &=& \phi_2 -\phi_1 = \gamma_1 \phi_1 \\
 & \vdots & \nonumber \\
 y_k & = & \phi_k-\phi_{k-1} =\phi_1 \prod_{j=1}^{k-1} \gamma_j
 \label{eqref12}
  \\
 & \vdots & \nonumber \\
 y_N & = & \phi_N-\phi_{N-1} = \phi_1 \prod_{j=1}^{N-1} \gamma_j
\end{eqnarray}
As usual, the empty product is one, $  \prod_{j=1}^{0} \gamma_j=1$.
Let us now calculate the sum over all $y_j$. This sum is a ``telescope-sum'' and simplifies to
\begin{eqnarray}
\sum_{k=1}^N y_k = \phi_1 -\underbrace{\phi_0}_0+\phi_2 -\phi_1 +\phi_3-\phi_2 + \ldots +\underbrace{\phi_N}_{1}-\phi_{N-1} =1.
\label{eqref13}
\end{eqnarray}
Using Eqs.~\eqref{eqref12} and \eqref{eqref13}, we can finally calculate $\phi_1$,
\begin{eqnarray}
1 = \sum_{k=1}^N y_k 
=  \sum_{k=1}^N \phi_1 \prod_{j=1}^{k-1} \gamma_j 
= \phi_1 \left( 1+ \sum_{k=1}^{N-1}  \prod_{j=1}^{k} \gamma_j  \right)
\end{eqnarray}
Hence, the fixation probability of a single $A$ individual, $\phi_1$, is given by 
\begin{equation}
\phi_1 = \frac{1}{1+ \sum_{k=1}^{N-1}  \prod_{j=1}^{k} \gamma_j}
\label{start}
\end{equation}
For $T_j^- = T_j^+$, we have $\gamma_j=1$. Hence, all products are simply one and we find $\phi_1 = 1/N$. This is the case of neutral selection, where all individuals have the same fitness. Any random walk in which the probability to move to the left or to the right is identical for the transient states leads to the same result. 

So far, we have focused on
the fixation probability of a single mutant in a resident population, $\phi_1$. In general, the fixation probability $\phi_i$ is given by
\begin{eqnarray}
\phi_i 
& = & \sum_{k=1}^{i} y_k \\
& = & \phi_1  \sum_{k=1}^{i}  \prod_{j=1}^{k-1} \gamma_j \\
& = & \phi_1  \left( 1+ \sum_{k=1}^{i-1}  \prod_{j=1}^{k} \gamma_j \right)\\
&=& \frac{1+ \sum_{k=1}^{i-1}  \prod_{j=1}^{k} \gamma_j}{1+ \sum_{k=1}^{N-1}  \prod_{j=1}^{k} \gamma_j}
\label{fixprob}
\end{eqnarray}
For neutral selection, we have $T^+_j=T^-_j$, which results in $\gamma_j=1$. In this case, the fixation probability reduces to $\phi_i = i/N$.

Formally, the fixation probability can be calculated in systems in which the replicator equation predicts coexistence, i.e.\ no fixation. However, it can also be shown that in these cases, the average time until fixation grows exponentially with increasing population size \cite{antal:2006aa} and increasing intensity of selection \cite{traulsen:2007cc}.

Often, the comparison between the probability that a single $A$ individual takes over a population of $N-1$ resident $B$ individuals, $\rho_A = \phi_1$, and the probability that a single $B$ individual takes over a population of $N-1$ resident $A$ individuals, $\rho_B$, is of interest in order to determine in which state the system spends more time
\cite{nowak:2006bo}. Formally, the probability $\rho_B$ is equal to the probability that $N-1$ individuals of type $A$ fail to take over a population in which there is just a single $B$ individual. Hence, we find 
\begin{eqnarray}
\rho_B &=& 1 - \phi_{N-1} \\
& = & 1 - \frac{1+ \sum_{k=1}^{N-2}  \prod_{j=1}^{k} \gamma_j}{1+ \sum_{k=1}^{N-1}  \prod_{j=1}^{k} \gamma_j} \\
& = & \frac{1+ \sum_{k=1}^{N-1}  \prod_{j=1}^{k} \gamma_j}{1+ \sum_{k=1}^{N-1}  \prod_{j=1}^{k} \gamma_j} 
- \frac{1+ \sum_{k=1}^{N-2}  \prod_{j=1}^{k} \gamma_j}{1+ \sum_{k=1}^{N-1}  \prod_{j=1}^{k} \gamma_j} \\
& = & \frac{\prod_{j=1}^{N-1} \gamma_j}{1+ \sum_{k=1}^{N-1}  \prod_{j=1}^{k} \gamma_j} \\
& = & \rho_A {\prod_{j=1}^{N-1} \gamma_j}
\end{eqnarray}
Therefore, the ratio of the two fixation probabilities is given by
$
\frac{\rho_B}{\rho_A} = {\prod_{j=1}^{N-1} \gamma_j}.
$
If this product is smaller than 1, we have $\rho_B<\rho_A$, if it is larger than 1, we have 
$\rho_B>\rho_A$. 
For small mutation rates, $\rho_B<\rho_A$ means that the system
spends more time in the $A$ state, because less invasion attempts are necessary to reach fixation by $A$. The $A$ mutants have a higher probability to reach fixation in a $B$ population compared to $B$ mutants in an $A$ population.

\subsection{Fixation times}

Another quantity of interest in evolutionary dynamics of finite population is the average time until fixation occurs \cite{ewens:2004qe}. For two strategies, three different fixation times are of interest: 
\begin{itemize}
\item[(i)] The average time $t_j$ until either one of the two absorbing states, $A$ or $B$, is reached when starting from state $j$. This is the unconditional fixation time.
For weak selection, this time increases with the distance 
between $j$ and the two absorbing boundaries.
\item[(ii)] The conditional fixation time $t_j^A$ specifies the average time it takes to reach the absorbing state $A$ when starting from state $j$, provided 
that $A$ is ultimately reached. 
The time $t_j^A$ increases with the distance between the states $j$ and $A$. 
If fixation of strategy $A$ is almost certain, 
$t_j^A$  is very similar to the unconditional fixation time $t$.
Of particular interest is $t_1^A$, or $t^A$ for short, which denotes the average fixation time of a single $A$ mutant in a resident $B$ population.
\item[(ii)] In analogy to $t^A_j$, $t^B_j$ represents the average time to reach the absorbing state $B$ when starting in state $j$ (i.e. with $j$ individuals of type $A$ and $N-j$ of type $B$), provided 
that $B$ is ultimately reached. 
$t^B_j$
increases with the distance 
between state $j$ and $B$. 
\end{itemize}

\paragraph{Unconditional fixation time.}
The unconditional average fixation time $t_j$, starting from state $j$, is determined by 
\begin{eqnarray}
t_j = 1+ T^-_j t_{j-1} +(1-T^-_j-T^+_j) t_j +  T^+_j t_{j+1}.
\label{timeq0}
\end{eqnarray} 
This equation can be rationalized as follows: In one time step (this results in the $1$), the process can either move to $j -1$, stay in $j$ or move to $j+1$. Then, the fixation time from that state matters. When we start in $0$ or in $N$, fixation has already occurred, thus $t_0=t_N=0$. 
Eq.~\eqref{timeq0} can be written as
\begin{eqnarray}
\underbrace{t_{j+1}-t_j }_{z_{j+1}}=  \gamma_j (\underbrace{ t_j-t_{j-1} }_{z_{j}})- \frac{1}{T_j^+}.
\label{timeq1}
\end{eqnarray} 
With the notation ${\gamma_j} = \frac{T_j^-}{T_j^+} $ from above, iteration yields
\begin{eqnarray}
z_1 &=& t_1 - t_0 =  t_1\\
z_2 &=& t_2 -t_1 = \gamma_1 t_1 -   \frac{1}{T_1^+}\\
z_3 &=& t_3 -t_2 = \gamma_2 \gamma_1 t_1 -   \frac{\gamma_2}{T_1^+} - \frac{1}{T_2^+}\\\
 & \vdots & \nonumber \\
 z_k & = & t_{k}-t_{k-1} =t_1 \prod_{m=1}^{k-1} \gamma_m - \sum_{l=1}^{k-1} \frac{1}{T_l^+} \prod_{m=l+1}^{k-1} \gamma_m
\end{eqnarray}
For the sum of the $z_k$, we find 
\begin{eqnarray}
\sum_{k=j+1}^N z_k = 
t_{j+1}-t_{j} + t_{j+2}-t_{j+1}  + \ldots + \underbrace{t_N}_{=0}-t_{N-1} =
- t_{j} 
\end{eqnarray}
In particular, we have for $j=1$
\begin{eqnarray}
t_1 = -\sum_{k=2}^N z_k 
= - t_1 \sum_{k=1}^{N-1} \prod_{m=1}^{k} \gamma_m + \sum_{k=1}^{N-1} \sum_{l=1}^{k} \frac{1}{T_l^+} \prod_{m=l+1}^{k} \gamma_m.
\end{eqnarray}
From this, the first fixation time, $t_1$, is obtained
\begin{eqnarray}
t_1 =\underbrace{ \frac{1}{1+\sum_{k=1}^{N-1} \prod_{j=1}^{k} \gamma_j }}_{\phi_1} 
\sum_{k=1}^{N-1} \sum_{l=1}^{k} \frac{1}{T_l^+} \prod_{j=l+1}^{k} \gamma_j.
\end{eqnarray}
Here, $\phi_1$ is the fixation probability given by Eq.~\eqref{start}.
Thus, the average unconditional fixation time for general $j$ is finally given by 
\begin{eqnarray}
t_{j} =-\sum_{k=j+1}^N z_k = - t_1\sum_{k=j}^{N-1}  \prod_{m=1}^{k} \gamma_m +\sum_{k=j}^{N-1} \sum_{l=1}^{k} \frac{1}{T_l^+} \prod_{m=l+1}^{k} \gamma_m.
\label{uncondavfixtime}
\end{eqnarray}
It is important to notice that the variance of these fixation times is usually high, depending on the 
population size, the game and the intensity of selection \cite{traulsen:2007cc,dingli:2007aa}. In particular for coexistence games, where the replicator dynamics predicts a stable coexistence of $A$ and $B$, the fixation times do not only diverge with the population size and the intensity of selection, but they represent averages of a very broad distribution  \cite{traulsen:2007cc}. 

\paragraph{Conditional fixation times.}
Given that the process reaches the absorbing state with $A$ individuals only, how long does this take
when starting in state $j$? 
To calculate this time $t_A^j$, we follow Antal and Scheuring \cite{antal:2006aa}. It is convenient to start from 
\begin{eqnarray}
\phi_j t^A_j = \phi_{j-1} T^-_j(t^A_{j-1}+1) +\phi_{j} (1-T^-_j-T^+_j)(t^A_{j}+1) + \phi_{j+1} T^+_j(t^A_{j+1}+1). \nonumber
\end{eqnarray}
Here, $\phi_j$ is the fixation probability of $j$ individuals of type $A$, see Eq.~\eqref{fixprob}. 
With the abbreviation $\theta_{j}^A = \phi_j t_j^A$, 
we can write this as
\begin{eqnarray}
\underbrace{\theta_{j+1}^A -\theta_{j}^A }_{w_{j+1}} = 
\underbrace{\theta_{j}^A -\theta_{j-1}^A }_{w_{j}} \frac{T_j^-}{T_j^+}
- \frac{\phi_j}{T_j^+}.
\label{timeeq2}
\end{eqnarray}
Eq.~\eqref{timeeq2}  has the same structure as Eq.~\eqref{timeq1}. Thus, we can use a similar iteration as above to obtain
\begin{eqnarray}
 w_k & = & \theta^A_{k}-\theta^A_{k-1} = \theta^A_1 \prod_{m=1}^{k-1} \gamma_m - \sum_{l=1}^{k-1} \frac{\phi_l}{T_l^+} \prod_{m=l+1}^{k-1} \gamma_m .
\end{eqnarray}
At the lower boundary, we have $\theta^A_0=0$, because $\phi_0=0$. We also have $\theta^A_N=0$ at the upper boundary, because $t^A_N=0$. 
Summing over $w_k$ leads to 
$\sum_{k=j+1}^N w_k = 
- \theta^A_{j} 
$.
In particular, for $j=1$, we obtain 
\begin{eqnarray}
t_1^A =
\sum_{k=1}^{N-1} \sum_{l=1}^{k} \frac{\phi_l}{T_l^+} \prod_{m=l+1}^{k} \gamma_m.
\label{t1aeq}
\end{eqnarray}
Often, this is the quantity of interest, because it corresponds to the average time it takes
for a single mutation to reach fixation in the population. 
For general $j$, we have
\begin{eqnarray}
t_j^A = - t_1^A  \frac{\phi_1}{\phi_j}
\sum_{k=j}^{N-1} \prod_{m=1}^{k} \gamma_m
+
\sum_{k=j}^{N-1} \sum_{l=1}^{k} \frac{\phi_l}{\phi_j}\frac{1}{T_l^+} \prod_{m=l+1}^{k} \gamma_m.
\end{eqnarray}
For $\phi_1=\phi_j=1$ (certain fixation of $A$), $t_j^A$ reduces to the unconditional fixation time, 
Eq.~\eqref{uncondavfixtime}.

For completeness, let us also calculate the average time $t_j^B$ until type $B$ reaches fixation in the population. Instead of Eq.~\eqref{timeeq2}, we now have
\begin{eqnarray}
\underbrace{\theta_{j}^B -\theta_{j-1}^B }_{v_{j}} =
\underbrace{\theta_{j+1}^B -\theta_{j}^B }_{v_{j+1}} 
\frac{1}{\gamma_j}
+ \frac{\tilde \phi_j}{T_j^-}, 
\end{eqnarray}
where $\tilde \phi_j=1-\phi_j$ is the fixation probability for reaching state $B$ 
and $\theta_{j}^B =\tilde \phi_j t_j^B$. Again, we have $\theta_{0}^B=\theta_{N}^B =0$.   
Now, we start our iteration from $j=N-1$,
\begin{eqnarray}
v_{N} &=& \theta^B_{N} -\theta^B_{N-1} =  -\theta^B_{N-1}\\
v_{N-1} &=&  \theta^B_{N-1} -\theta^B_{N-2} = 
- \theta^B_{N-1}  \frac{1}{ \gamma_{N-1}} +   \frac{\tilde \phi_{N-1}}{T_{N-1}^-}\\
 & \vdots & \nonumber \\
 v_{N-k} & = &\theta^B_{N-k} -\theta^B_{N-k-1} =
 -\theta_{N-1}^B \prod_{m=1}^{k} \frac{1}{\gamma_{N-m}} 
 +
  \sum_{l=1}^{k} \frac{\tilde \phi_{N-l}}{T_{N-l}^-} \prod_{m=l+1}^{k} \frac{1}{\gamma_{N-m}}
  \nonumber.
\end{eqnarray}
Summation yields $\sum_{k=N-j}^{N-1} v_{N-k}=\theta^B_j$. From $j=N-1$, we find for the fixation time of a single $B$ mutant
\begin{eqnarray}
t_{N-1}^B = \sum_{k=1}^{N-1} \sum_{l=1}^k  \frac{\tilde \phi_{N-l}}{T_{N-l}^-}\prod_{m=l+1}^{k} \frac{1}{\gamma_{N-m}}.
\end{eqnarray}
Of course, this quantity can also obtained from Eq.~\eqref{t1aeq} by symmetry arguments. 
Finally, we obtain for the time until fixation of  $B$ when 
starting from an arbitrary number of $B$ individuals
\begin{eqnarray}
t_j^B = 
- t_{N-1}^B \frac{\tilde \phi_{N-1}}{\tilde \phi_{j}} \sum_{k=N-j}^{N-1} \prod_{m=1}^k \frac{1}{\gamma_{N-m}}
+
 \sum_{k=N-j}^{N-1}  \sum_{l=1}^{k} \frac{\tilde \phi_{N-l}}{\tilde \phi_{j}}  \frac{1}{T_{N-l}^-} \prod_{m=l+1}^{k} \frac{1}{\gamma_{N-m}}.
\end{eqnarray}
This formulation is valid for general birth-death processes. In addition to the 
fixation probabilities, the two types of fixation
times are of particular interest 
to characterize the evolutionary process
because they represent 
global quantities that include information on all transition probabilities.

\newpage

\subsection{The Moran process and weak selection}

As a specific example, let us consider the 
frequency dependent Moran process
\cite{nowak:2004pw,taylor:2004wv}. Apart from calculating the above quantities for a specific case, our goal is to find simpler expressions for the fixation probabilities under weak selection. 
In this case, we are close to neutral selection, which seems to be a biologically relevant limit \cite{crow:1970ck,ohta:2002aa}.

First, we need to
specify the payoffs of the two types $A$ and $B$.  
The payoffs of A and B individuals, $\pi_A$ and $\pi_B$, are given by
\begin{eqnarray}
\pi_A &=& \frac{j-1}{N-1} a + \frac{N-j}{N-1} b 
\label{eq29}
\\
\pi_B &=& \frac{j}{N-1} c + \frac{N-j-1}{N-1} d .
\label{eq30}
\end{eqnarray}
Here, we have excluded self interactions, i.e. in a population of $j$ individuals of type $A$, each one of them interacts with $j-1$ others of its type. 
Fitness is assumed to be a linear combination of background fitness (which we set to 1) and the payoff,
\begin{eqnarray}
f_A &=& 1-w+w \pi_A \\
f_B &=& 1-w+w \pi_B .
\end{eqnarray}
The transition probabilities are
\begin{eqnarray}
T^+_j &=& \frac{j \; f_A}{j \; f_A + (N-j) f_B} \frac{N-j}{N} \\
T^-_j &=& \frac{(N-j) \; f_B}{j \; f_A + (N-j) f_B} \frac{j}{N}  .
\end{eqnarray}
For the ratio of the transition probabilities, we have
\begin{equation}
\gamma_j = \frac{T^-_j}{T^+_j} = \frac{f_B}{f_A} = \frac{1-w+w \pi_B}{1-w+w \pi_A} .
\end{equation}

Let us now consider the limit of weak selection, i.e.\ $w \ll 1$, and derive an approximation for the fixation probability $\phi_1$.
For weak selection, $\gamma_j$ simplifies to
\begin{eqnarray}
\gamma_j = \frac{1-w+w \pi_B}{1-w+w \pi_A} 
\approx  1-w (\pi_A - \pi_B) .
\label{gammaapprox}
\end{eqnarray}
The product in Eq.~(\ref{start}) can then be simplified to 
\begin{eqnarray}
\prod_{j=1}^k \gamma_j & \approx & \prod_{j=1}^k (1-w(\pi_A-\pi_B)) \approx  1-w  \sum_{j=1}^k (\pi_A-\pi_B).
\end{eqnarray}
Next, we introduce a new notation for $\pi_A-\pi_B$. From Eqs.~(\ref{eq29}) and (\ref{eq30}), we find
\begin{equation}
\label{uv}
\pi_A-\pi_B = \underbrace{\frac{a-b-c+d}{N-1}}_u j + \underbrace{ \frac{-a+bN-dN+d}{N-1}}_v .
\end{equation}
With this we can solve the sum over the payoff difference,
\begin{eqnarray}
\sum_{j=1}^k (\pi_A-\pi_B) = \sum_{j=1}^k (u \; j +v) 
 =  u \frac{(k+1)k}{2} + v k
  =  \frac{u}{2}k^2 + \left( \frac{u}{2} +v \right) k.
 \label{end}
\end{eqnarray}
Now we have derived a simple form for $\prod_{j=1}^k \gamma_j$ under weak selection. Let us analyze the ratio of fixation probabilities:
\begin{eqnarray}
\label{thetadef}
\frac{\rho_B}{\rho_A} & =& {\prod_{j=1}^{N-1} \gamma_j} 
 \approx  1-w \sum_{j=1}^{N-1} \left(\pi_A -\pi_B \right) \\ \nonumber
& = & 1-w \left[ \frac{u}{2}(N-1) +  \frac{u}{2} +v  \right] (N-1) \\ \nonumber
& = & 1-\frac{w}{2} \underbrace{\left[ (a-b-c+d)(N-1) - a-b-c+3d+(2b-2d)N \right]}_{\Xi}.
\end{eqnarray}
For $\Xi>0$ we have $\rho_A>\rho_B$. For large populations, $N \gg 1$, we can approximate
\begin{eqnarray}
0< \Xi \approx 
N(a+b-c-d),
\end{eqnarray}
which is equivalent to 
\begin{eqnarray}
x^{\ast} =
\frac{d-b}{a-b-c+d} <\frac{1}{2}
\end{eqnarray}
Hence, $\rho_A>\rho_B$ is equivalent to ${x^{\ast}} <\frac{1}{2}$. As we have discussed above,  this condition indicates risk dominance and establishes a relation to fixation probabilities: 
For weak selection, strategies with higher fixation probabilities have greater basins of attraction. 

Inserting Eq.~(\ref{end}) into Eq.~(\ref{start}), we obtain an approximation for the fixation probability of a single $A$ individual:
\begin{eqnarray}
\phi_1 &=& \frac{1}{1+ \sum_{k=1}^{N-1}  \prod_{j=1}^{k} \gamma_j} 
 \approx  \frac{1}{1+ \sum_{k=1}^{N-1}  \left[1-w \left(  \frac{u}{2} k^2 + \left( \frac{u}{2} +v \right) k \right) \right]}.
\end{eqnarray}
Using 
$\sum_{k=1}^{N-1} k = {N (N-1)}/{2}$
and 
$\sum_{k=1}^{N-1} k^2 = {N (N-1)(2N-1)}/{6}$,
the fixation probability becomes 
\begin{eqnarray}
\phi_1 
& \approx &\frac{1}{N- w  u \frac{N (N-1)(2N-1)}{12} - w \left( \frac{u}{2} +v \right) \frac{N(N-1)}{2}  } \\
\nonumber
& = &\frac{1}{N} + \frac{w}{4N}  \underbrace{ \left[ (a-b-c+d) \frac{2N-1}{3} -a-b-c+3d +(2b-2d)N \right] }_{\Gamma}  
\label{fixprobwsmoran}  
\end{eqnarray}
The same fixation probability under weak selection is found for a large variety of processes \cite{imhof:2006aa,traulsen:2005hp,lessard:2007aa}. For general $i$, we have
\begin{eqnarray}
\phi_i 
 \approx 
\frac{i}{N} + N w \frac{N-i}{N} \frac{i}{N}  
\left( 
\frac{a-b-c+d}{6(N-1)}(N+i) + \frac{-a+bN-dN+d}{2(N-1)}
\right).
\label{fixprobwsmorangeneral}  
\end{eqnarray}

Let us now compare the fixation probability $\phi_1$ to the result for neutral selection, $w=0$. 
This leads to the so called $1/3$-rule. Neutral selection means that we have no selective forces and that only randomness determines fixation. 
In this case, we have $\phi_1=1/N$. Since we are only interested if the fixation probability is larger or smaller than $1/N$, we have to consider only the sign of $\Gamma$. If $\Gamma>0$, the fixation probability is larger than $1/N$. 
For large $N$ this condition reduces to
\begin{equation}
\frac{a-b-c+d}{3} +b-d >0.
\end{equation}
This condition is equivalent to 
\begin{equation}
x^{\ast}=
 \frac{d-b}{a-b-c+d} < \frac{1}{3}.
\end{equation}
The 1/3-rule states that ``In a coordination game, the fixation probability of a strategy under weak selection is larger than $1/N$, if the unstable fixed point is closer than $1/3$ to the strategy to be replaced". 
The intuitive justification of this rule goes back to the fact that during invasion, a single invader will interact on average $1/3$ with its own type and $2/3$ with the other type \cite{ohtsuki:2007aa}. 

If we increase the advantage of strategy $A$ in such a coordination game systematically (e.g. by increasing the payoff against itself), and hence shifting the mixed equilibrium $x^\ast$ to lower values, the following scenarios occur \cite{nowak:2004pw}: 
\begin{center}
\begin{tabular}{l l}
$x^\ast>2/3$ 	&  $A$ is disadvantageous and $B$ is advantageous \\
			& ($\rho_A<1/N$ and $\rho_B>1/N$) \\
$2/3>x^\ast>1/2$ & $B$ is risk dominant, but both $A$ and $B$ 
are
disadvantageous \\
			&  ($\rho_A<\rho_B$, $\rho_A<1/N$ and $\rho_B < 1/N$) \\
$1/2>x^\ast>1/3$ & $A$ becomes risk dominant, but both $A$ and $B$ remain
still
  \\
			& disadvantageous ($\rho_A>\rho_B$, $\rho_A<1/N$ and $\rho_B < 1/N$) \\
$x^\ast<1/3$ &  $A$ is advantageous and $B$ is disadvantageous \\
			& ($\rho_A>1/N$ and $\rho_B<1/N$) \\
\end{tabular}
\end{center}

Interestingly, a analogous condition as the $1/3$-rule also holds for coexistence games. In this case, the stable interior fixed point has to be closer than $1/3$ to the strategy that is reached during fixation. In other words, 
``In a coexistence game, the fixation probability of a strategy under weak selection is larger than $1/N$, if the stable fixed point is further than $2/3$ to the strategy to be replaced". 
However, fixation probabilities are only of limited  interest here, as fixation times become very large for large populations. 
The fixation times under weak selection are discussed in \cite{altrock:2008aa}.

\subsection{The Fermi process}
\label{egfs}

The Moran process leads to simple analytical results under weak selection
but no similar simplifications are possible for higher selection strengths. 
In contrast, a pairwise comparison process with $p$ given by the Fermi function of the payoff difference (see above) admits simple analytical results for any intensity of selection. 
The transition probabilities are
\begin{eqnarray}
T^{\pm}_j &=& \frac{j}{N} \frac{N-j}{N} \frac{1}{1+e^{\mp w(\pi_A - \pi_B)}}.
\end{eqnarray}
The analytical accessibility of this process is based on a very simple ratio of these transition probabilities,
\begin{equation}
\gamma_j = \frac{T^-_j}{T^+_j} = e^{-w (\pi_A- \pi_B)}.
\end{equation}
In the weak selection limit, $w \ll 1$, we recover Eq.~\eqref{gammaapprox}.
Thus, the fixation probabilities are identical to the Moran process 
and the $1/3$ rule remains valid. 
For arbitrary values of $w$, 
let us 
return to the ratio of fixation probabilities:
\begin{eqnarray}
\frac{\rho_B}{\rho_A} & =& {\prod_{j=1}^{N-1} \gamma_j} 
=  \exp \left[- w \sum_{j=1}^{N-1} \left(\pi_A -\pi_B \right) \right] 
=  \exp \left[- \frac{w}{2} {\Xi} \right].
\end{eqnarray}
Here, $\Xi$ is defined as in Eq.~\eqref{thetadef}. Again
we have $\rho_A>\rho_B$ for $\Theta>0$. 
For large $N$, we find again that $\rho_A>\rho_B$ is equivalent to ${x^{\ast}} <\frac{1}{2}$. But now, the relation between the fixation probabilities and risk dominance is valid for arbitrary intensities of selection, not only for weak selection. 

The expressions for fixation probabilities simplify, because the products over $\gamma_j$ reduce to sums that can be solved exactly. One special case is determined by  frequency independence of the payoff difference
$a-c=b-d$. This case has been termed ``equal gains from switching'', because switching from strategy $B$ to $A$ leads to the same payoff change, 
irrespective of the opponents move \cite{nowak:1990aa}. In this special case, even the outer sum in Eq.~\eqref{fixprob} can be solved exactly for any $w$. We find
\begin{equation}
\phi_i = \frac{1-e^{-w\, v\, i}  }{1-e^{-w\, v\, N}  }.
 \label{pd}
 \end{equation}
This result is identical to the fixation probability of $k$ individuals with fixed relative fitness $r=e^{w v}$ \cite{crow:1970ck,ewens:2004qe}. Thus, a game with equal gains from switching has the same fixation properties as constant selection in frequency independent settings with fixed fitness values.
Since the Fermi process only depends on payoff differences, this is no surprise. But it also shows that
properties of constant selection apply not only for the Moran process under weak selection but for other processes as well.

\begin{figure}[t]
\def\capfrac{1}
\begin{center}
\includegraphics[width=0.8\textwidth]{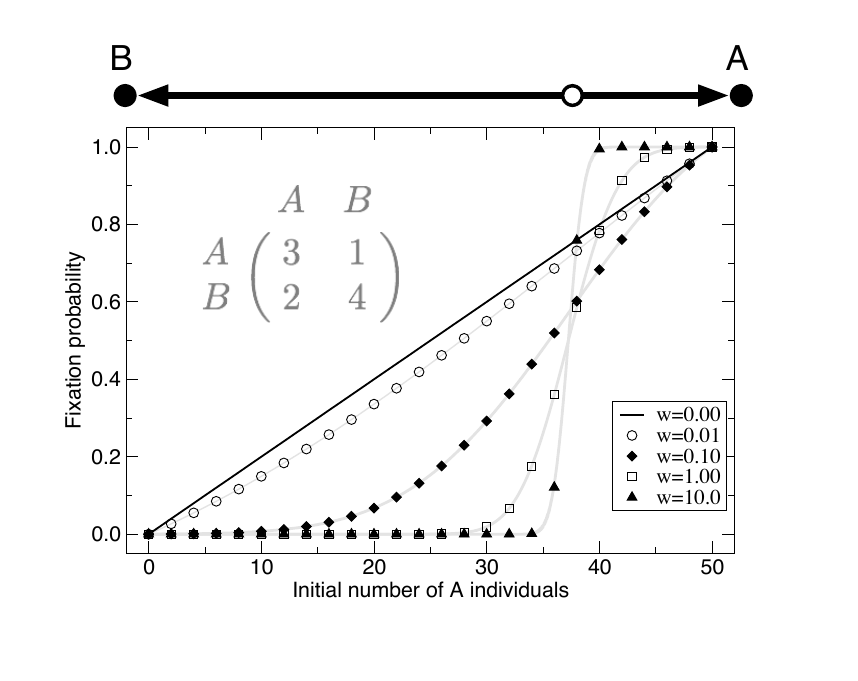}
\end{center}
\Caption{
Fixation probabilities for a coordination game with bistability (diagram on top). For neutral selection, $w=0$, the fixation probabilities are a linear function of the initial number of $A$ individuals
(solid line). With increasing intensity of selection $w$, the fixation probabilities of the pairwise comparison process (with $p$ given by the Fermi function) converge to a step function with a discontinuity at the unstable fixed point, $x^{\ast} = 37.5$. Symbols represent the exact expression and lines the continuous approximation Eq.~\eqref{fixationerf} (payoff matrix given in the figure, $N=50$).  }
\label{fig2}
\end{figure}

For general payoffs, we can approximate the outer sum in Eq.~\eqref{fixprob} over $k$ by an integral, 
$\sum_{k=1}^i \ldots 
 \approx
\int
_1^i \ldots dk$ and arrive at \cite{traulsen:2006bb}
 \begin{equation}
\phi_k= \frac{{\rm erf}\left[ Q_k\right]-
{\rm erf}\left[ Q_0\right]}
{{\rm erf}\left[Q_N \right]-
{\rm erf}\left[ Q_0\right]}.
\label{fixationerf}
\end{equation}
Here, 
${\rm erf}(x)=\frac{2}{\sqrt{\pi}}\int_0^x dy\, e^{-y^2}$ is the error function
and 
$Q_k=\sqrt{\frac{w(N-1)}{ 2u}} \left(k  u + v \right) $ 
with $u, v$ as in Eq.~(\ref{uv})
\cite{traulsen:2006bb}. 
The result is valid for $ u \neq 0$
and in the limit of $u \to 0$, it reduces to Eq.~\eqref{pd}. 
In the weak selection limit, $w\to 0$, 
Eqs.\ (\ref{fixationerf}) and (\ref{pd}) 
recover the neutral selection result $\phi_k =k/N$.
Numerical simulations of the fixation probabilities agree very well with this approximation and hold even for small populations where the approximation of sums by integrals becomes inadequate (see Fig.~\ref{fig2}). 

The Fermi process covers all intensities of selection and leads to strong selection results that are outside the realm of the standard Moran process. 
The closed expressions allow to derive approximations for the fixation probabilities under weak and under strong selection. 
As for the Moran process,  the time to fixation grows exponentially with $N$ for games with internal Nash equilibria, such that fixation will practically never occur. 
Here, it also grows exponentially with the intensity of selection $w$.

\section{From finite to infinite populations (and back again)}

So far, we have introduced two descriptions for evolutionary game dynamics: The deterministic replicator dynamics for large populations and stochastic evolutionary game dynamics in finite populations. Here, we discuss how both dynamics are related to each other. 
In the limit of large $N$, this is accomplished by performing a Kramers-Moyal expansion of the Master equation 
\cite{gardiner:1985bv,kampen:1997xg}: 
\begin{eqnarray}
 P^{\tau+1}(j) - P^{\tau}(j)
&=&
\hphantom{+}
P^{\tau}(j-1)  T^+_{j-1}+  P^{\tau} (j+1) T^- _{j+1}
 \nonumber \\
&  & -P^{\tau}(j)  T^-_{j}
-P^{\tau}(j)  T^+_{j},
\end{eqnarray}
where $P_j^{\tau}$ denotes the probability to be in state $j$ at time $\tau$. To consider large $N$, we introduce the notation 
$x=j/N$, $t=\tau/N$ and the 
probability density $\rho(x,t) = N\, P^{\tau}(j)$. For the transition probabilities, we replace $T^{\pm}_j \to  T^{\pm}(x)$. 
This
yields
\begin{eqnarray}
 \rho\left(x,t \! + \! \frac{1}{N} \right) 
 -  \rho\left(x,t\right) \! \! \!
&= & \! \!
\! \rho \! \left(x-\frac{1}{N},t\right)   T^+(x-\frac{1}{N})
\nonumber
 + \rho \! \left(x+\frac{1}{N},t\right)  T^- (x+\frac{1}{N})
 \nonumber \\  &-& \! \! \!
 \rho\left(x,t\right)  T^- (x)
 -\rho\left(x,t\right)  T^+ (x).
\label{ma2}
\end{eqnarray}
For $N \gg 1$, the probability densities and
the transition probabilities are expanded in a Taylor series at $x$ and $t$. 
More specific, we have 
\begin{eqnarray}
\rho\left(x ,t +  \frac{1}{N} \right) 
& \approx & 
\rho\left(x,t\right) 
+  \frac{\partial}{\partial t}\rho\left(x,t\right) \frac{1}{N} 
\\
\rho\left(x \pm  \frac{1}{N},t\right) 
&\approx &
\rho\left(x,t\right) 
\pm  \frac{\partial}{\partial x}\rho\left(x,t\right) \frac{1}{N} 
+ \frac{\partial^2}{\partial x^2}\rho \left(x,t\right) \frac{1}{2N^2}
\end{eqnarray}
and
\begin{eqnarray}
T^{\pm}(x  \pm   \frac{1}{N}) 
\approx 
T^{\pm}(x) 
\pm  \frac{\partial}{\partial x}T^{\pm} (x) \frac{1}{N} 
+ \frac{\partial^2}{\partial x^2}T^{\pm} (x) \frac{1}{2N^2} .
\end{eqnarray}
Let us now look at the terms depending on their order in $1/N$. 
The terms independent of $1/N$ cancel on both sides of Eq.~(\ref{ma2}). 
The first non-vanishing term is of order $1/N$. On the left hand side, we have the term
$ \frac{\partial}{\partial t}\rho\left(x,t\right)$ and on the right hand side, we have 
\begin{eqnarray}
\nonumber
& & -\rho(x,t) \frac{\partial}{\partial x}T^+(x)
+\rho(x,t) \frac{\partial}{\partial x}T^- (x)
-T^+ (x) \frac{\partial}{\partial x} \rho(x,t)
+T^- (x) \frac{\partial}{\partial x} \rho(x,t) \\
& & = - \frac{\partial}{\partial x} \left[ T^+(x) - T^-(x) \right] \rho(x,t) .
\end{eqnarray}
This term describes the average motion of the system. 
In physics, it is called the drift term but in biology, it is referred to as selection term.  
Next, we consider terms of the order $1/N^2$. On the right hand side, we have
\begin{eqnarray}
\nonumber
&& \hphantom{+}\left( \frac{\partial}{\partial x} \rho(x,t) \right) \left( \frac{\partial}{\partial x} T^+(x) \right) 
+ \frac{1}{2}\rho(x,t) \frac{\partial^2}{\partial x^2} T^+(x) + \frac{1}{2} T^+(x)  \frac{\partial^2}{\partial x^2}\rho(x,t)
\\
\nonumber
&& +\left( \frac{\partial}{\partial x} \rho(x,t) \right) \left( \frac{\partial}{\partial x} T^-(x) \right) 
+ \frac{1}{2}\rho(x,t) \frac{\partial^2}{\partial x^2} T^-(x) + \frac{1}{2} T^-(x)  \frac{\partial^2}{\partial x^2}\rho(x,t)  \\
& &= \frac{1}{2} \frac{\partial^2}{\partial x^2} \left[ T^+(x) + T^-(x) \right] \rho(x,t) .
\end{eqnarray}
This second term, called diffusion in physics, leads to a widening of the probability distribution in the course of time. In biology, it is called genetic or neutral drift, which can be a source of confusion. In the following, higher order terms will be neglected. Thus, we can approximate Eq.~(\ref{ma2}) by
\begin{equation}
\frac{\partial}{\partial t}\rho\left(x,t\right) = - \frac{\partial}{\partial x} \underbrace{\left[ T^+(x) - T^-(x) \right]}_{a(x)} \rho(x,t) +  \frac{1}{2} \frac{\partial^2}{\partial x^2} \underbrace{ \frac{ T^+(x) + T^-(x) }{N}}_{b^2(x)} \rho(x,t)
\end{equation} 
This is the Fokker-Planck equation of the system, which describes the deterministic time evolution of a probability distribution. 

Equivalently, one can describe the process by a stochastic differential equation that generates a single trajectory. If the noise is microscopically uncorrelated, as in our case, the It{\^o} calculus has to applied \cite{gardiner:1985bv}. In this framework, the Fokker-Planck equation above corresponds to the stochastic differential equation 
\begin{equation}
\dot x = a(x) + b(x) \xi,
\end{equation}
where $\xi$ is uncorrelated Gaussian noise, $a(x)$ is the drift term (selection in biology). In general, the diffusion term $b(x)$ (genetic drift in biology) depends not only on the composition of the population, but also on the payoffs. However, for many processes, the payoff dependence vanishes. In particular for weak selection, $b(x)$ is independent of the payoffs. Note that the noise is multiplicative and that the drift term vanishes at the boundaries $x=0$ and $x=1$, which is important to avoid that these boundaries are crossed. For additive noise, processes leading to $x<0$ or $x>1$ have to be excluded artificially \cite{traulsen:2004iq} 

\begin{figure}[t]
\def\capfrac{1}
\begin{center}
\includegraphics[width=1.0\textwidth]{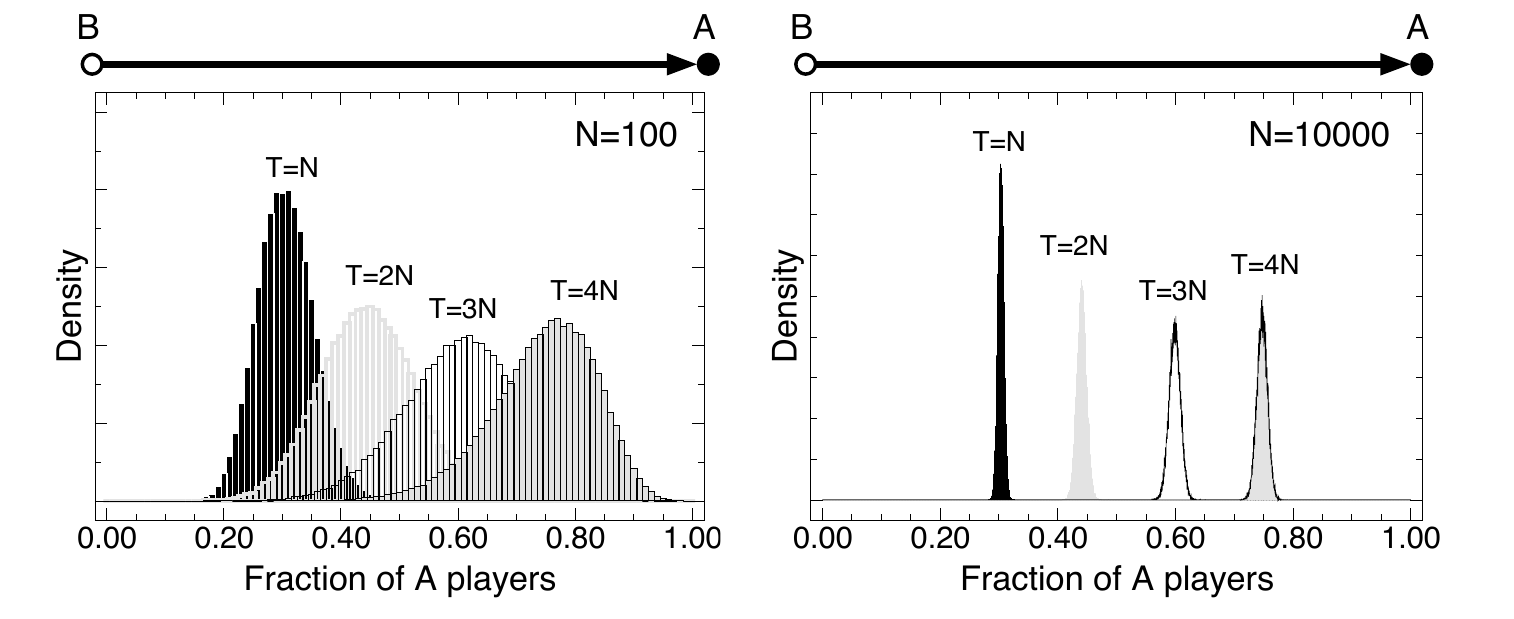}
\end{center}
\Caption{
With increasing population size, the dynamics converges to the replicator equation in a game where strategy $A$ dominates
$B$ (diagram on top). We start 
with $20\%$ individuals of type $A$ and 
depict the strategy distribution after $T=N, 2N, 3N$ and $4N$ time steps
for a small population of size $N=100$ (left) as well as a large population with $N=10000$ (right). Selection shifts the distribution towards higher fractions of $A$ players, while the stochasticity of the evolutionary process widens the distribution. For small populations this  quickly increases the variance, whereas for large populations the effect is barely visible and the distribution remains narrow.
Parameters: payoff matrix $a=2$, $b=5$, $c=1$, $d=3$, simulations with pairwise comparison based on the Fermi function with $w=1$, averages over $10^5$ realizations.
}
\label{fig3}
\end{figure}

For $N \to \infty$, we have $b(x) \to 0$ and only the term $a(x)$ determines the dynamics. This case reduces to the deterministic differential equation,
\begin{eqnarray}
\dot x = T^+(x) - T^-(x)
\end{eqnarray}
and recovers the replicator equation \cite{traulsen:2005hp} (see Fig.~\ref{fig3}). The same procedure also works for more than two strategies, although in this case, the mathematics
is more tedious \cite{traulsen:2006hp}. A similar procedure can also be applied to spatially extended systems, where a stochastic partial differential equation is obtained \cite{reichenbach:2007aa,reichenbach:2007bb}.
Note that we have only considered the limit $N \to \infty$, keeping everything else fixed. To perform a thermodynamical limit, the intensity of selection $w$ has to be scaled with $N$ \cite{chalub:2006cc}.

The approximation of the process for large $N$ can be used to address the evolutionary dynamics in large populations under weak selection. For example, we can 
verify the $1/3$-rule. Starting from the Fokker-Planck equation, the probability of fixation can be written in terms of the drift and diffusion coefficients as 
\begin{equation}
\phi_j  = \frac{S(j)}{S(N)} 
\hspace{0.5cm}{\rm where} \hspace{0.3cm}
S(j)=\int_0^{j/N} \exp \left[ -2 \int_0^y \frac{ a(z)}{b^2(z)} dz \right] dy,
\label{fixation}
\end{equation}
see \cite{ewens:2004qe,gardiner:1985bv,traulsen:2006ab}. 
For neutral selection, we have $a(z)=0$ and thus $\phi_j  = \frac{j}{N}$.  
In general, the comparison of the full fixation probability to neutral selection has to be done numerically.
In the limit of weak selection, $w\ll 1$ the $1/3$-rule is recovered when comparing the fixation probability of a small group of mutants to neutral selection \cite{traulsen:2006ab}.
More precisely, the $1/3$ rule is 
obtained for $N w \ll 1$. For $N w \gg 1$, the result expected from the replicator equation is recovered: A disadvantageous 
rare strategy will 
never reach fixation with a significant probability. 

\section{Applications}
Complementing the general theory of evolutionary game dynamics we now turn to the two most important applications in evolutionary biology and across behavioral sciences: the problem of cooperation and the maintenance of biodiversity. In game theory, cooperation refers to behavioral actions that benefit others at some cost to the actor. Thus, groups of cooperating individuals fare better than groups of non-cooperating defectors. However, each individual faces the temptation to defect and free-ride on the benefits produced by others. This generates a conflict of interest between the individuals and the group, which characterizes social dilemmas \cite{dawes:1980aa,hauert:2006fd,doebeli:2005aa}. Social dilemmas are abundant in nature. They represent a recurrent theme ranging from defense formations in musk oxen to defend their young against wolves \cite{hamilton:1971jt}, sentinel behavior in meerkats \cite{clutton-brock:1999aa}, predator inspection in fish \cite{milinski:1987ju,pitcher:1992aa}, grooming in baboons \cite{saunders:1988aa,stammbach:1982aa}, protein production in phages \cite{turner:1999hp,turner:2003hp} to microorganisms producing extra-cellular products such as enzymes in yeast \cite{greig:2004aa}, biofilms \cite{rainey:2003an} or antibiotic resistance \cite{neu:1992aa}, to name only a few prominent examples. However, social dilemmas also occurred on evolutionary scales and life could not have unfolded without the repeated incorporation of lower level units into higher levels entities. Every resolution of a social dilemma marks a major transition in evolutionary history: the formation of chromosomes out of replicating DNA, the transition from unicellular to multicellular organisms or from individuals to societies all require cooperation \cite{maynard-smith:1995bo}. In human interactions, social dilemmas are equally abundant in terms of social security, health care and pension plans but even more importantly when it comes to the preservation of natural resources from local to global scales, including drinking water, clean air, fisheries and climate \cite{hardin:1968mm,milinski:2006lr,Milinski:2008lr}.

The viability of ecological systems is determined by the biodiversity, which includes species, habitat and genetic diversity \cite{tilman:nature06,hector:nature07,storch:07}. Species co-existence is promoted by non-hierarchical, cyclic interactions where $R$ beats $S$ beats $P$ beats $R$, just as in the children's game Rock-Scissors Paper. This occurs in bacterial strains of \emph{E. coli}, where a neutral strain is eliminated by a toxin producing strain, which is then outgrown by an immune but non-toxic strain, which is in turn outgrown by the neutral strain and so on. Further examples of cyclic dominance hierarchies include mating strategies in lizards \cite{sinervo:1996le,sinervo:2006aa} or competition for space in coral reef invertebrates \cite{jackson:1975aa} and links to the problem of cooperation if participation in social dilemmas is voluntary rather than compulsory \cite{hauert:2002te,hauert:2002in,semmann:2003he} (see Sect.~\ref{loner}).

\subsection{The Prisoner's Dilemma}

Let us begin with the Prisoner's Dilemma, which has a long tradition as a mathematical metaphor to analyze the problem of cooperation \cite{axelrod:1981yo,axelrod:1984yo}. 
In the Prisoner's Dilemma, two players can choose between cooperation and defection. A cooperative act costs $c>0$, but leads to a benefit $b>c$ for the other player. Thus, the highest payoff, $b$, is obtained when only the partner is cooperating. In this case, the partner obtains $-c$. Mutual cooperation leads to $b-c$ and mutual defection a zero payoff. The game is characterized by the payoff matrix 
\begin{equation}
\bordermatrix{
  & C & D \cr
C & b-c & -c \cr
D & b & 0 \cr}.
\end{equation}
The Prisoner's Dilemma represents the most stringent form of a social dilemma because the
strategy $D$ dominates strategy $C$.
No matter what the opponent does, one should defect
since $b>b-c$ and $0>-c$.
Unilateral deviation from mutual defection decreases the payoff and hence mutual defection represents the only Nash equilibrium \cite{nash:1950ef,holt:2004mm}
but mutual cooperation corresponds to the social optimum ($b-c>0$). 

Note that the parametrization in terms of costs and benefits represents the most intuitive and mathematically convenient form of the Prisoner's Dilemma. However, it is equally important to note that this reflects a special case because
the sum of the diagonal elements equals the sum of the non-diagonal elements of the payoff matrix. In other words, the game is an example of ``equal gains from switching'' (see section \ref{egfs}). This property leads to a payoff difference between cooperators and defectors, $\pi_C-\pi_D=-c$, that is independent of the fraction of cooperators $x_C$.
In this special case the replicator dynamics reads
\begin{equation}
\dot x_C = - x_C(1-x_C) c
\end{equation}
and can be solved exactly:
$x_C(t) = x_C(0) \left[ 
x_C(0)+ \left(1-x_C(0) \right) e^{+c t}
\right]^{-1} .
$
The fraction of cooperators $x_C$ is always decreasing and converges to the only stable fixed point $x_C=0$. Cooperators are doomed and disappear. 

In finite populations and under weak selection, $w \ll 1$, we find, in agreement with Eq.~\eqref{fixprobwsmoran}, for the fixation probability of $i$ cooperators in a population of $N-i$ defectors 
\begin{equation}
\phi_i = \frac{i}{N}- \frac{i}{N} \frac{N-i}{N}\left(c+ \frac{b}{N-1}\right) \frac{N w}{2}< \frac{i}{N}.
\end{equation}
Since $\phi_i< \frac{i}{N}$, cooperators are at a disadvantage compared to neutral mutants. 
Similarly, for strong selection, we find from the Fermi process in the limit $w \to \infty$  the fixation probabilities $\phi_i = \delta_{i,N}$. In other words, cooperation cannot evolve from individual selection alone. 

The stark contrast between theoretical predictions of the Prisoner's Dilemma and the observed abundance of cooperation in nature calls for explanations. Over the last decades, a number of mechanisms capable of promoting cooperation in biological and social systems have been proposed \cite{nowak:2006pw,doebeli:2005aa}. Among related individuals, cooperators may thrive due to kin selection \cite{hamilton:1964bo} and competition among groups may promote cooperation through group selection \cite{wilson:1975sg,wilson:1994gs,fletcher:2004bv,traulsen:2006aa}. Conditional behavioral rules that strategically respond to previous encounters in repeated interactions or adjust their behavior according to an individuals' reputation in non-repeated settings can establish cooperation through direct \cite{trivers:1971hp} or indirect reciprocity \cite{nowak:1998is}. Local interactions in structured populations support cooperation in the Prisoner's Dilemma \cite{nowak:1992pw,ohtsuki:2006na} but not necessarily in other social dilemmas \cite{hauert:2004bo}. Finally, extensions of the strategy space that enable individuals to abstain from social dilemmas and render participation voluntary or options to punish non-cooperating interaction partners both support cooperation \cite{hauert:2002te,hauert:2007aa} (see Sects.~\ref{loner},~\ref{peer}).

\subsection{Rock-Paper-Scissors}
\label{rps}

Rock-paper-scissors games are the simplest example of cyclic dominance, where any strategy can be beaten by another one:
Rock crushes scissors, scissors cut paper and paper wraps rock. 
This simple game not only entertains children (and adults \cite{rps}) but equally serves as a mathematical metaphor to investigate the dynamics and maintenance of biodiversity \cite{reichenbach:2007aa,szolnoki:2004aa,szabo:2007jt}.
The game can be characterized by the matrix
\begin{equation}
\label{Mmatrix}
	\bordermatrix{
		  & R & P & S \cr
		R & 0 & -s & +1 \cr
		P & +1 & 0 & -s \cr
		S & -s & +1 & 0 \cr
		}.
\end{equation}
This parametrization assumes symmetric strategies
but the payoff for winning ($+1$) is not necessarily equal to the payoff for losing ($-s<0$). 
For the standard choice $s=1$, we have a zero-sum game
-- one player's gain is the other ones' loss.

\begin{figure}[t]
\def\capfrac{1}
\begin{center}
\includegraphics[width=1.0\textwidth]{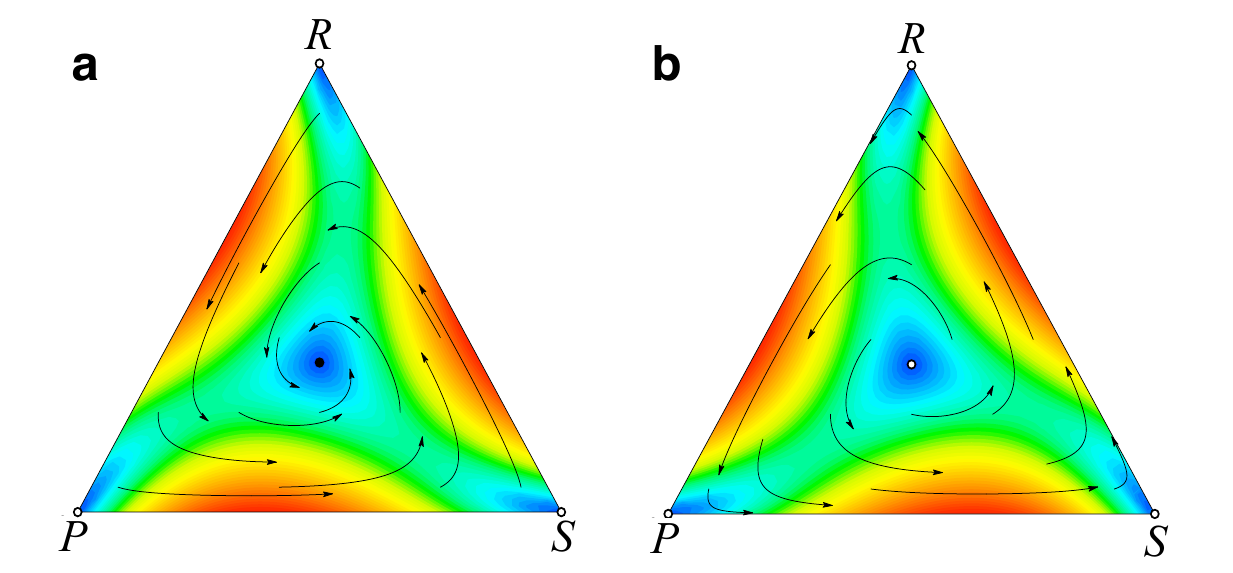}
\end{center}
\Caption{
Depending on the determinant of the payoff matrix $d$, two qualitatively different scenarios of cyclic dominance are possible in the game of Rock-Paper-Scissors. 
(a) If the determinant $d$ is positive, the interior fixed point is asymptotically stable and all orbits converge to this point.
(b) If the determinant $d$ is negative, the interior fixed point is unstable and the system
approaches a heteroclinic cycle along the boundary of the simplex
(the figure is produced with Bill Sandholm's Dynamo package \cite{sandholm:2007dy}). 
}
\label{rpsfig}
\end{figure}

Apart from the three trivial homogeneous equilibria, the replicator dynamics admits a non-trivial
equilibrium at $\boldsymbol{x^{\ast}} = (x_R,x_P,x_S) =(\frac{1}{3},\frac{1}{3},\frac{1}{3})$.
The dynamics of the system is determined by the Lyapunov function
\begin{equation}
H=-x_R x_P x_S .
\label{constantofmotion}
\end{equation}
If the determinant of the payoff matrix~(\ref{Mmatrix}), $d=1-s^3$, is positive, then 
$\frac{\partial H}{\partial t} <0$ and 
the interior fixed point ${\boldsymbol x^{\ast}}$ is
asymptotically stable. 
For $d<0$, the fixed point ${\boldsymbol x^{\ast}}$ is unstable and 
the system
approaches a heteroclinic cycle along the boundary of the simplex $S_3$.
Finally, for the zero-sum game with $s=1$, the function
(\ref{constantofmotion})
is a constant of motion, and the system infinitely oscillates
around ${\boldsymbol x^{\ast}}$, see Fig.~\ref{rpsfig}.

In finite populations, 
the dynamics is stochastic but characteristic features can be derived 
from the average drift towards the internal fixed point \cite{claussen:2008aa}. 
For large $N$, the average drift of the Moran process computed over the entire simplex 
is given by
\begin{eqnarray}
\langle \Delta H \rangle_N
\label{morandrift}
= \frac{1}{20N^2}
-\frac{1-s}{420N}\left(
\frac{1}{2}
-\frac{1}{N}
\right) w.
\end{eqnarray}
For $N \to \infty$, the replicator equations lead to $\Delta H=0$ for
zero-sum games ($s=1$) as well as for neutral selection ($w=0$). In finite populations, we obtain $\langle \Delta H \rangle_N =\frac{1}{20N^2}>0$ in both cases instead. 
For $s=1$, stochasticity turns the neutrally stable, closed orbits of infinite populations into an unstable system by introducing a drift towards the heteroclinic cycle.
Only in the limit $N \to \infty$, the neutrally stable orbits characterized by $\Delta H=0$ are recovered from the Moran process or other forms of finite population dynamics. 
However, even in the infinite system it is not trivial to numerically determine these closed orbits of the replicator equation and the numerical integrator has to be chosen with great care \cite{hofbauer:1996mm}.

For $s<1$, the sign of 
$\langle \Delta H \rangle_N$ depends on the intensity of selection $w$, the payoff parameter $s$ as well as on the population size $N$. This leads to a critical population size
\begin{equation}
N_c = 2+\frac{42}{w(1-s)}.
\end{equation}
Since $42$ is the answer to life, the universe, and everything, this result is not surprising \cite{adams:1979aa}. 
For $N<N_c$, the system cycles towards the boundaries and fixation is expected to be fast. For $N>N_c$,  
the system converges towards   ${\boldsymbol x^{\ast}}$ on average. 

This game is an example of a system that changes its qualitative dynamics if the population falls below a certain threshold. Such thresholds are often found in other evolutionary games in finite populations \cite{taylor:2004wv,nowak:2004pw,claussen:2007aa,traulsen:2005hp}.

\newpage

\subsection{\label{loner}Voluntary Public Goods Games}

So far, we focussed on interactions among pairs of individuals. Now, we turn to Public Goods Games \cite{kagel:1997aa} in which a group of $M$ players interact. Public Goods Games represent a generalization of pairwise Prisoner's Dilemma interactions to interaction groups of arbitrary size $M$ \cite{hauert:2003aa}.

In typical Public Goods experiments, $M$ individuals have the opportunity to cooperate and invest a fixed amount $c$ into a common pool or to defect and invest nothing. The total investment is multiplied by a factor $r$ and distributed equally among all participants -- irrespective of whether they contributed or not. Thus, every invested unit returns $r c/M$ units to the investor (as well as to all other participants). If $r < M$ then rational players withhold their investments because they are costly -- but if all participants reason in this manner, no investments are made and the group foregoes the benefits of the public good. In contrast, had everybody cooperated, they would have been better off with a return of $(r-1)c$. Again we have a social dilemma and the Nash equilibrium is different from the social optimum. However, if $r > M$ then investments have a positive net return, and rational players will invest in the public good \cite{hauert:2006fd}.

In infinite populations with a fraction $x$ cooperators and $y$ defectors ($x+y=1$) the average payoff of defectors is $\pi_D = (M-1)x r c/M$ and of cooperators it is $\pi_C=\pi_D-(1-r/M)c$. Thus, for $r<M$ cooperators decline and eventually disappear because $\pi_C<\pi_D$. However, for $r>M$ cooperators dominate ($\pi_C>\pi_D$) and eventually displace defectors. In this case cooperation evolves as a by-product \cite{connor:1995aa}.

\begin{figure}[t]
\def\capfrac{1}
\begin{center}
\includegraphics[width=1.0\textwidth]{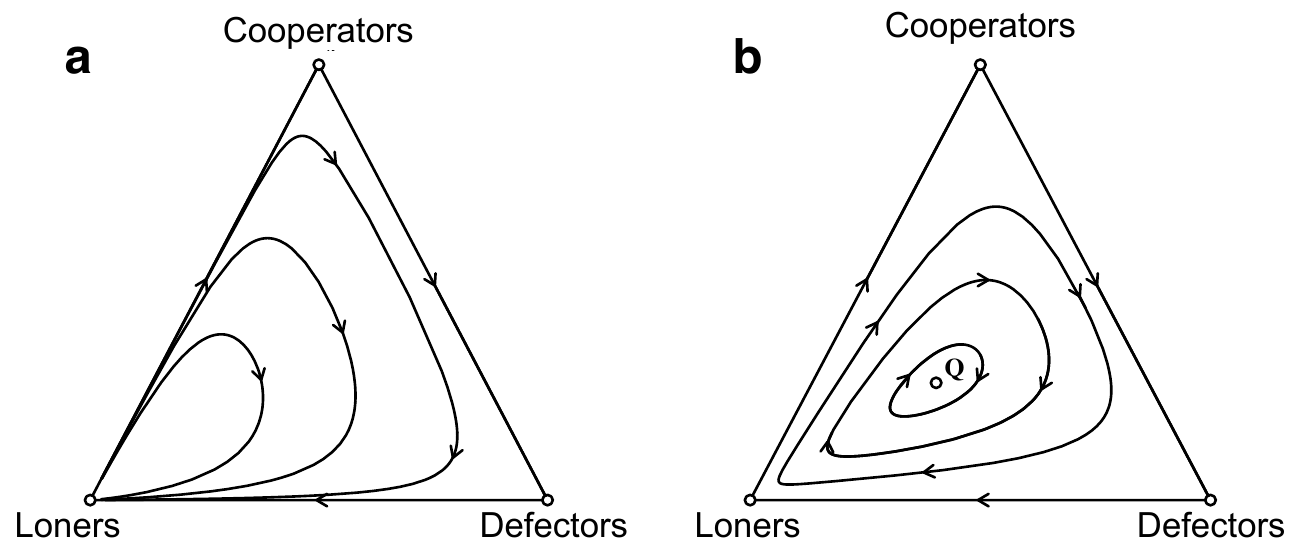}
\end{center}
\Caption{
Dynamics in voluntary Public Goods Games in which cooperators are dominated by defectors, defectors by loners and loners by cooperators.
(a) For $r \leq 2$, only brief intermittent bursts of cooperation are observed before the
system settles again in the loner state. 
(b) For $r>2$, a neutrally stable fixed point $\bf Q$ appears, which is surrounded by closed orbits. The fraction of cooperators performs stable oscillations (parameters: (a) $c=1$, $r=1.8$ and $\sigma=0.5$. (b)  $c=1$,  $r=3$ and $\sigma=1$).
}
\label{lonerfig}
\end{figure}

The above analysis is based on the assumption of compulsory participation in public goods interactions but what happens if participation is voluntary and risk averse individuals may choose to abstain? This situation can be modeled by introducing a third strategic type, the non-participating loners \cite{hauert:2002in,hauert:2002te}. Loners obtain a fixed payoff $\sigma$, which lies between the payoff in groups consisting solely of cooperators, $(r-1)c$, and the payoff in groups of defectors (which is zero). Interaction groups are randomly formed by sampling $M$ individuals from an infinite population with a fraction $x$ cooperators, $y$ defectors and $z$ loners. Thus, the effective interaction group size $S$ of the Public Goods Game decreases with increasing abundance of loners. If only a single individual participates in the Public Goods interaction, the game does not take place and the individual is forced to act as a loner. The average payoffs of defectors, cooperators and loners are given by \cite{hauert:2002in}:
\begin{eqnarray}
\pi_D & = & 
\sigma z^{M-1} + r \; c \frac{x}{1-z} \left(1-\frac{1-z^M}{M(1-z)} \right) \nonumber\\ 
\pi_C & = & \pi_D
-\left(1+(r-1)z^{N-1}-\frac rM \frac{1-z^M}{1-z}\right)c \nonumber\\
\pi_L & = & \sigma
\end{eqnarray}
Note that with few loners, $z\to 0$, defectors dominate cooperators, $\pi_D>\pi_C$, but for $z\to 1$, cooperators outperform defectors, $\pi_C>\pi_D$. This generates a Rock-Scissors-Paper type cyclic dominance among the three strategies: if cooperators abound it pays to defect and if defection is common it is best to abstain but this reduces the effective interaction group size $S$ until eventually $r>S$ holds and cooperators thrive, which in turn increases $S$, restores the social dilemma and the cycle continues, see Fig.~\ref{lonerfig}b. By definition $S\geq 2$ must hold because a single participant ($S=1$) is unable to sustain the public good. Therefore, the above reasoning requires $r>2$. In this case, a heteroclinic orbit along the boundary of the simplex $S_3$ reflects the cyclic dominance of the three strategies and in the interior a neutrally stable fixed point $\bf Q$ exists, which is surrounded by closed orbits such that the system exhibits stable periodic oscillations of the strategy frequencies, see Fig.~\ref{lonerfig}b. For $r\leq 2$, the heteroclinic orbit still exists but the interior is filled with homoclinic orbits of the state with all loners, $z=1$. Therefore, only brief intermittent bursts of cooperation are observed before the system returns to the loner state, see Fig.~\ref{lonerfig}a.

In order to allow for a compact analytical description of the stochastic dynamics in finite populations, we assume small mutation rates $\mu$
such that a strategy reaches fixation or extinction much faster than mutants are produced \cite{imhof:2005oz,fudenberg:2006fu,hauert:2007aa}. 
This is fulfilled if $\mu$ is much smaller than the inverse of the squared population size. Thus, the system is usually homogeneous and only occasionally switches from one homogeneous state to another. The dynamics is essentially reduced to a stochastic process along the boundaries of the discretized simplex $S_3$ and 
we can approximate the system by an embedded Markov chain on the pure states of the system, $C$, $D$ and $L$. 
The average time the system spends in each state depends on the updating process, the intensity of selection as well as on the game parameters.
For simplicity, let us consider imitation dynamics (c.f. Eq.~(\ref{fermieq}) in the limit of strong selection, $w\to\infty$).
In this case, 
a single defector takes over a cooperator population with probability $1$.
Similarly, the probability that a single loner takes over a population of defectors is also $1$. Finally, a loner population is taken over by cooperators with probability $1/2$. This is because the first cooperator is neutral (no Public Goods games take place) and disappears with the same probability as it gets imitated. However, as soon as there are two cooperators, they have an advantage and eliminate the loners with certainty. This leads to the transition matrix among the three homogeneous states:
\begin{equation}
\label{cdltrans}
\bordermatrix{
  	& C 					& D				& 	L			 	\cr
C	& \frac{1}{2} 			& 0				& \frac{1}{4} 		\cr
D 	& \frac{1}{2}			& \frac{1}{2}		& 0		 		\cr
L	& 0					& \frac{1}{2}		& \frac{3}{4}		\cr}.
\end{equation}
Note that the system stays e.g. in state $C$ with probability $1/2$ because the probability that the mutant is an unsuccessful loner is $1/2$. Under imitation dynamics the transition matrix is parameter independent as long as the cyclic dominance applies (which only requires $r<M$ and $0<\sigma<r-1$).
The stationary distribution ${\boldsymbol P}$ is given by the eigenvector corresponding to the eigenvalue $1$ of this stochastic matrix, ${\boldsymbol P}=(P_C,P_D,P_L)=(\frac{1}{4},\frac{1}{4},\frac{1}{2})$.
Thus, the system spends 50\% of the time in the loner state and 25\% in the cooperator and and defector states, respectively. In compulsory interactions, i.e. in the absence of loners,
the system would spend essentially all the time in the defector state. Interactive Java simulations of this system can be found online \cite{vlabs:2008}.

In summary, voluntary participation in Public Goods Games provides an escape hatch out of states of mutual defection. This maintains cooperation, but fails to stabilize it. 
The cyclic dynamics of voluntary Public Goods Games has been confirmed in behavioral experiments \cite{semmann:2003he}.
In the next section, we demonstrate that the time spent in cooperative states can be vastly extended by introducing opportunities to punish defectors.

\subsection{\label{peer}Punishment}

Punishment is ubiquitous in nature ranging from bacterial colonies to human societies \cite{clutton-brock:nature95,fehr:2002bv,foster:2000le,gurerk:2006jb,sigmund:2007oz,dreber:2008lr}. In game theoretical terms, punishment is defined as behavioral actions that impose a fine $\beta$ on a co-player at some cost $\gamma<\beta$ to the actor. Punishment enables cooperators to retaliate against non-cooperating defectors and therefore seems capable of stabilizing cooperation. However, this raises a second-order social dilemma because non-punishing cooperators outperform those that do punish. Thus, in an evolving population, mild cooperators undermine the punishers' efforts and pave the way for the successful invasion of defectors. Moreover, if punishers are rare, they suffer tremendous costs from punishing left and right, while inflicting little harm on the defecting members of the population and hence it remains unclear how punishment behavior could have gained a foothold in a population. Colman aptly summarizes this by stating: ``we seem to have replaced the problem of explaining cooperation with that of explaining altruistic punishment'' \cite{colman:2006aa}.

In the previous section we demonstrated that voluntary participation in Public Goods Games promotes cooperation without being able to stabilize it. Here we extend this setting by introducing a fourth strategic type, the punishers. Punishers are individuals that cooperate but, in addition, punish defectors. The combination of volunteering an punishment was originally proposed by Fowler \cite{fowler:2005aa}, but it turns out that the replicator dynamics in infinite populations is unsuitable to draw clear-cut conclusions because the system is bi-stable (i.e. the evolutionary end state depends on the initial configuration) and also structurally unstable (i.e. the evolutionary dynamics may undergo significant changes upon introducing stochastic components) \cite{brandt:2006aa}.

\begin{figure}[t]
\def\capfrac{1}
\begin{center}
\includegraphics[width=1.0\textwidth]{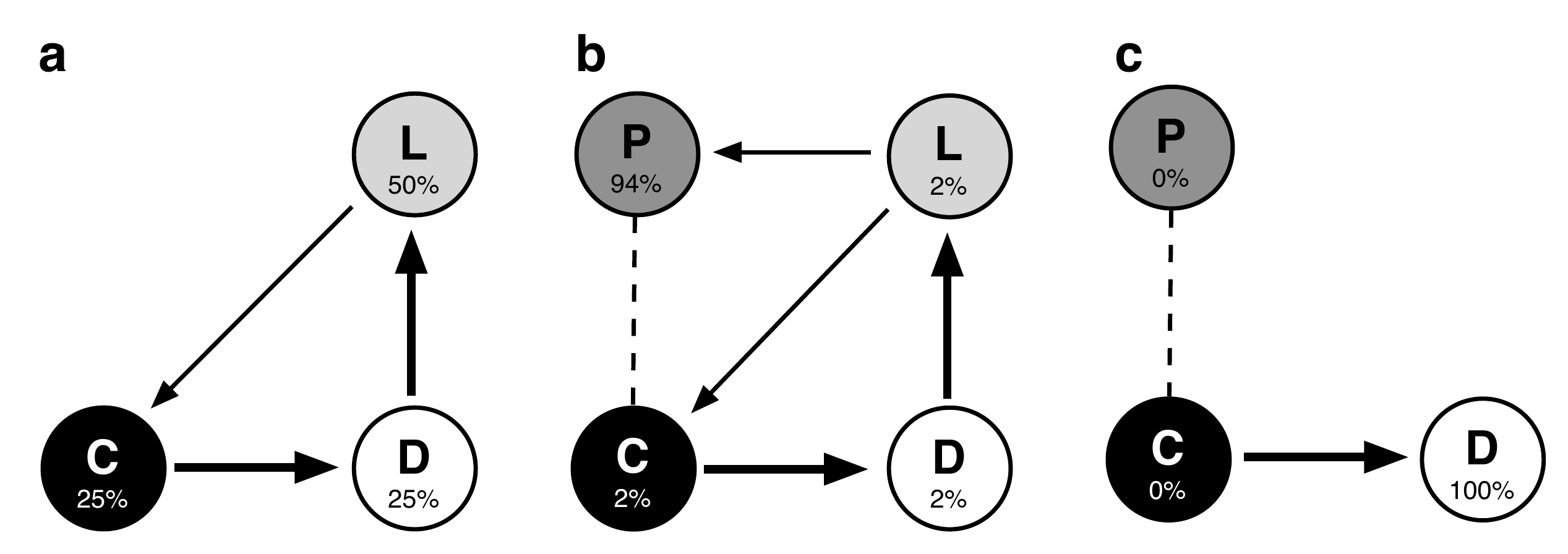}
\end{center}
\Caption{
For small mutation rates and strong selection, the stochastic dynamics of finite populations can be approximated by an embedded Markov chain on the pure states of the system. 
(a) For the voluntary Public Goods Game, a cyclic dominance between cooperators, defectors and loners emerges. We find that the system spends 50\% in the Loner state and 25 \% in the cooperator and defector states, respectively. This result is independent of all parameters.
(b) In voluntary Public Goods Games with punishment, punishers can invade the loner state. The system then spends a long time in the punisher state because only cooperators can invade and only through neutral drift. Once the system is in the cooperator state it is prone to invasion by defectors and may undergo several cooperator-defector-loner-cycles before punishment is re-established. In the long run, punishers dominate the population most of the time.
(c) In compulsory Public Goods Games with punishment, the ultimate outcome is defection. Even when starting in the punisher state, cooperators can invade by neutral drift and
once defectors take over, the system has reached its evolutionary end state 
(population size $N=92$).}
\label{smarties}
\end{figure}

In the following, we consider the stochastic dynamics of volunteering and punishment in Public Goods Games in finite populations \cite{hauert:2007aa,hauert:2008bb}. As in the case without punishment, we take the limit of small mutation rates and strong selection. Again, this implies that the population is homogeneous most of the time, i.e. in states $C, D, L$ or $P$. An occasional mutant will have taken over the population or disappeared before the next mutation arises. Therefore, the dynamics is determined by a Markov chain based on the different transition probabilities. In the limit of strong selection, the derivation of these transition probabilities is particularly simple: 
(i) in the cooperator state $C$, a single defector is advantageous and takes over the population with probability $1$.  
Loners are disadvantageous and cannot invade. 
Punishers can invade through neutral drift with probability $\frac{1}{N}$. 
(ii) in the defector state $D$, a mutant cooperator or punisher is disadvantageous and disappears. 
Loners are advantageous and take over with probability $1$.
(iii) in the loner state $L$, the dynamics is neutral if a mutant of any type arises. With probability $1/2$ another individual adopts the mutant strategy (or the mutant disappears with the same probability). A pair of mutant cooperators or punishers is advantageous and takes over but a pair of defectors is disadvantageous and at least one disappears. (iv) in the punisher state $P$, both defectors and loners are disadvantageous. The former because they are punished and the latter because they do not take advantage of the common good. However, cooperators obtain the same payoff as punishers and can take over through neutral drift with probability $\frac{1}{N}$.
This yields the transition matrix 
\begin{equation}
\bordermatrix{
  	& C 					& D	& 	L		& P	 	\cr
C	& \frac{2}{3}-\frac{1}{3N} 	& 0		& \frac{1}{6} 	& \frac{1}{3N}	\cr
D 	& \frac{1}{3}			& \frac{2}{3}	& 0		& 0 		\cr
L	& 0					& \frac{1}{3}	& \frac{2}{3}& 0		\cr
P	& \frac{1}{3N}			& 0		& \frac{1}{6}	& 1-\frac{1}{3N} 		\cr} ,
\end{equation}
which is again independent of the interaction parameters (c.f. Eq.~(\ref{cdltrans})) and only depends on the population size $N$ because of the neutral transitions between cooperators and punishers.
The stationary distribution becomes 
\begin{equation}
{\boldsymbol P}=(P_C,P_D,P_L,P_P)=
\left(
\frac{2}{8+N},\frac{2}{8+N},\frac{2}{8+N},\frac{2+N}{8+N}
\right).
\end{equation} 
Thus, for large $N$, the system spends almost all the time in the punisher state. The reason is that the transition leading away from the punisher state is neutral and thus very slow compared to all other transitions. The punisher state is the only pure Nash equilibrium, but it is not a strict Nash equilibrium, as cooperators are equally well off. Despite the vanishing time the system spends in the loners' state, voluntary participation plays a pivotal role for cooperation and punishment because it provides recurrent opportunities for establishing social norms based on punishing non-cooperating defectors, see Fig.~\ref{smarties}.

In contrast, in
compulsory Public Goods Games, i.e.\ in the absence of loners, 
the cyclic dominance of $C$, $D$, and $L$ is missing and once cooperation breaks down, it cannot get re-established, see Fig.~\ref{smarties}. 
Interestingly, punishment emerges only if the participation in Public Goods Games is voluntary. This conclusion nicely relates to recent experiments, where individuals can choose whether to join a Public Goods Game with or without punishment, they voluntarily commit themselves to the sanctioning rules \cite{gurerk:2006jb}.
For interactive simulations of volunteering and punishment, see \cite{vlabs:2008}.

This approach to punishment, which is most common in behavioral economics, raises moral concerns because enforcing cooperation through peer-punishment means that individuals take the law into their own hands, but mob law is clearly socially unacceptable. 

\section{Concluding remarks}
This review mainly focusses on the particularly insightful class of $2 \times 2$ games.
We have analyzed the stochastic dynamics of evolutionary games in 
finite populations, as it is, for example, described by the Moran process.  
The connection to the traditional description of evolutionary games by the
deterministic replicator equation is established 
through approximations in the limit of large populations.

As applications and extensions of the theoretical framework, we provide brief excursions into interactions with more strategic options, as in the case of Rock-Scissors-Paper games, which are relevant in the context of biodiversity, as well as to interactions among larger groups of individuals to address the problem of cooperation in the Prisoner's Dilemma and in Public Goods Games with voluntary participation and/or punishment opportunities.
Other important games that have not been covered here include the 
Snowdrift Game \cite{doebeli:2004bo,hauert:2004bo}, where cooperation is not 
strictly dominated by defection, and the Minority Game, which turned into a paradigm 
for simplified market models \cite{challet:1997uw,challet:2004bo,coolen:2005bo}. 

Further important directions of the field include 
spatial games, which have recently been summarized in
an excellent review by Szab{\'o} and F{\'a}th \cite{Szabo:2007aa}, as well as recent advances in
ecological games with variable population densities \cite{hauert:2006ha,hauert:tpb08}
and for
games with continuous strategy spaces \cite{killingback:procb99,doebeli:2004bo} 
based on
adaptive dynamics \cite{dieckmann:jmb96,geritz:evolecol98,metz:96}.

Over the past few years, the tremendous progress of our understanding of evolutionary dynamics can be largely attributed to numerous insights gained from stochastic dynamics in finite populations as well as from considering the analytically accessible limits of weak selection and rare mutations. The often counter intuitive results 
are inaccessible from traditional approaches based on replicator equations.

Stochastic evolutionary game dynamics is more difficult to handle analytically as compared to deterministic approaches such as the replicator dynamics. Nonetheless, it is a very powerful tool, because it 
implements a natural source of noise that actually renders the results more robust. For example, in bistable systems it allows to calculate transition rates between the different
states
rather than concluding that the dynamics depends on the initial conditions. Most importantly, any real population is subject to noise and incorporating 
such essential features into game theoretical scenarios significantly improves their relevance in modeling the real world.

\subsection*{Acknowledgements}
We thank W.H. Sandholm for providing the software for Fig.~4. 
A.T.\ acknowledges support by the ``Deutsche Akademie der Naturforscher Leopoldina'' (Grant No.\ BMBF-LPD 9901/8-134) and the Emmy-Noether program of the DFG. 
C.H.\ is supported by the John Templeton Foundation.

\setlength{\bibindent}{6mm} 
\renewcommand{\bibname}{References}

\end{document}